\documentclass[prb,twocolumn,showpacs,preprintnumbers,amsmath,amssymb]{revtex4}

\usepackage{graphicx}
\usepackage{dcolumn}
\usepackage{bm}
\renewcommand{\vec}[1]{\boldsymbol{#1}}

\newcommand{\sinc}{\text{sinc}}
\newcommand{\secondtoggle}[1]{\tilde{\bar{#1}}}
\newcommand{\secondaverage}[1]{\bar{\bar{#1}}}
\newcommand{\mansfield}[1]{[\![#1]\!]}

\def\Ib{{\bf I}}
\def\rb{{\bf r}}

\def\<{\langle}
\def\>{\rangle}

\begin{document}

\preprint{}

\title{Efficient decoupling and recoupling in solid state NMR for quantum computation}

\author{Fumiko Yamaguchi}
\altaffiliation[Also at ]{Physics Department, Keio University, Yokohama, Kanagawa 223-8522, Japan}
\author{Thaddeus D. Ladd}
\author{Cyrus P. Master}
\author{Yoshihisa Yamamoto}
\altaffiliation[Also at ]{National Institute of Informatics, Hitotsubashi, Chiyoda-ku, Tokyo 101-8430, Japan}%
\affiliation{%
E. L. Ginzton Laboratory, Stanford University, Stanford, CA 94305,
USA
}%

\author{Navin Khaneja}
\affiliation{Division of Applied Sciences, Harvard University,
Cambridge, MA 02138, USA
}%

\date{\today}

\begin{abstract}
A scheme for decoupling and selectively recoupling large networks
of dipolar-coupled spins is proposed.  The scheme relies on a
combination of broadband, decoupling pulse sequences applied to
all the nuclear spins with a band-selective pulse sequence for
single spin rotations or recoupling. The evolution-time overhead
required for selective coupling is independent of the number of
spins, subject to time-scale constraints, for which we discuss the
feasibility. This scheme may improve the scalability of
solid-state-NMR quantum computing architectures.
\end{abstract}

\pacs{%
82.56.Jn  
03.67.Lx, 
03.67.Pp. 
}
\maketitle

\section{Introduction}
Nuclear spins have been proposed as quantum bits (qubits) for
quantum computation, due to their good isolation from the
environment.\cite{Gershenfeld97,Cory97} With the aid of the
established technology of nuclear magnetic resonance (NMR) and the
use of ``pseudo-pure" states,\cite{Gershenfeld97,Cory97,Knill98}
liquid NMR at room temperature has allowed the largest quantum
computers to date.\cite{Vandersypen01}

In liquid NMR, a solution is used in which each solute molecule
has several nuclei, each nucleus having a different Larmor
frequency in the presence of a magnetic field.  This allows
selective addressing of the nuclei through application of resonant
radio-frequency (RF) fields at the corresponding frequencies. Each
synthesized molecule in the aqueous solution undergoes the same
logic operations.

Within each molecule, logic operations are performed by successive
application of single-bit rotations (by
RF pulse sequences) and multi-qubit conditional operations (by
time-evolution in the presence of coupling among the qubits). The
design of such logic gates relies on the assumption that the
coupling among the specific qubits can be set at will to {\it on}
(coupled) for construction of a logic gate, and {\it off}
(decoupled) for absence of a logic gate. A systematic scheme for
this has been developed,\cite{Leung00} but requires sequences of
duration $O(n)$
as the number of qubits $n$ increases, resulting in a clock-speed
that decreases as the computer-size increases.\cite{Ladd01}
Because this slow-down is only linear in $n$, and because some
quantum algorithms afford exponential speed-up, the scheme is
efficient in principle. However, truly scalable operation will
require quantum fault tolerance,\cite{faulttolerant} which
requires logic gates to be sufficiently accurate and fast in
comparison to decoherence times to surpass the fault-tolerance
threshold.  Faster logic gates are therefore of crucial importance
for future NMR quantum computers.

As an alternative to a liquid, the use of dipolar-coupled nuclear
spins in a crystalline solid \cite{Yamaguchi99,Ladd02} possesses
certain merits for scalability, such as long coherence
times,\cite{Ladd04} the ability to polarize the nuclear spins by
optical pumping of electron spins,\cite{Verhulst04} and more
sensitive methods of nuclear-spin detection.\cite{Ladd02,fu04}
However, it introduces additional complexity to the design of the
pulse sequences needed for logic operations. In these proposed
schemes, a magnetic field gradient is introduced to differentiate
qubit ensembles, by analogy to the chemical shift in liquid-NMR.
Unlike liquid-NMR, however, nuclei within a single qubit ensemble
are dipole-coupled. The complexity of pulse sequence design arises
from the additional burden of constantly decoupling these
intra-ensemble nuclear spins, which have the same Larmor
frequency, while also decoupling and recoupling nuclear spins with
distinct Larmor frequencies.

In this article, we propose a scheme that addresses both of these
issues: the linear slowing of logic operations and the
complication of added couplings in a crystalline state.  The
required ingredients are that all spins must be coupled by
strictly dipolar (or pseudo-dipolar) couplings in a large magnetic
field, the field gradient or chemical shift distinguishing these
qubits must be very large, and the couplings between different
qubits must be much stronger than couplings within an ensemble. As
previously discussed,\cite{Ladd02} this last condition may be
achieved in a geometry where nuclei are placed in one-dimensional
atomic chains, with a field gradient parallel to the chains.
Multiple, well-separated chains form the weakly coupled ensemble.
Other geometries, such as well-separated ensembles of planar
monolayers with two-dimensional gradients,\cite{goldman00} are
possible as well.

\section{Description of the Spin System}
In the presence of a static magnetic field along the $z$-axis, the
Hamiltonian consists of the Zeeman energy of the nuclear spins and
the coupling between nuclear spins by the dipole-dipole
interaction. The Zeeman term is written
\begin{equation}
H_{\rm Z} = -\gamma\hbar \sum_k B_k^{\phantom{z}} I_{k}^z =
-\sum_k\hbar\omega_{k}^{\phantom{z}}I_{k}^z, \label{eq:zeeman}
\end{equation}
where $\gamma$ is the gyromagnetic ratio of the nuclear spin and
$B_k$ is the magnetic field at the location of nuclear spin $k$.
Due to the field gradient, the magnetic fields $B_k$ and therefore
the Larmor frequencies $\omega_k$'s of the nuclear spins differ.
We denote the gradient-induced resonant frequency separation
between adjacent nuclei as $\omega_{k+1}-\omega_k=\delta\omega.$
The average Larmor frequency, which we notate $\omega_0$, is much
faster than all other frequencies in the system.

The dipole-dipole interaction between two nuclear spins $k$ and
$l$ separated by the vector $\rb_{kl}$ is given
by,\cite{Abragam61}
\begin{eqnarray}
H_{\rm D}
    = \frac{\mu_0}{4\pi}\frac{\hbar^2\gamma^2}{r_{kl}^3}\left\{
    \Ib_{k}\cdot\Ib_{l} - 3\frac{(\Ib_{k}\cdot\rb_{kl})(\Ib_{l}\cdot\rb_{kl})}{r_{kl}^2}
\right\}.
\end{eqnarray}
In the presence of a large applied field, the terms similar to
$I_k^z I_l^x$ or $I_k^x I_l^y$ oscillate at frequencies near
$\omega_0$ or $2\omega_0$, respectively.  These nonsecular terms
contribute negligibly to the evolution in the timescale under
consideration.\cite{Abragam61} With this approximation, the dipole
coupling for a pair of nuclear spins $k$ and $l$ takes the form
\begin{equation}\label{eq:couplingz}
H_{\rm D}^z =
D_{kl}^{\phantom{z}}\left(2I^z_kI^z_l-I^x_kI^x_l-I^y_kI^y_l\right),
\end{equation}
where
\begin{equation}
D_{kl}=\frac{\mu_0}{4\pi}\gamma^2\hbar^2\frac{1-3\cos^2\theta_{kl}}{2r_{kl}^3},
\end{equation}
and $\theta_{kl}$ is the angle between $\rb_{kl}$ and the
$z$-axis. 
The dipole coupling strength $D_{kl}$ between neighboring nuclei
is typically several hundred hertz, but can be smaller depending
on the geometry of the system.

The nuclear spins are manipulated by an oscillating RF magnetic
field $B_1$ in the plane perpendicular to the $z$-axis.
In the following, a ``broadband" pulse (or ``hard" pulse) with
Rabi frequency $\gamma B_1=\Omega_{\rm RF}$ has a pulse duration
$\sim 1/\Omega_{\rm RF}$ much smaller than $1/n\delta\omega$. A
``selective" pulse (or ``soft" pulse) with Rabi frequency
$\omega_{\rm RF}$ has a pulse duration much longer than
$1/\delta\omega$ in order to selectively rotate spins of a single
Larmor frequency $\omega_k$. Nearest-neighbor spins receive an
erroneous rotation of order $\sinc(\pi\delta\omega/2\omega_{\rm
RF})$; these erroneous rotations can be corrected by pulse shaping
techniques discussed elsewhere.\cite{Steffen00}

The couplings between nuclear spins and ``broadband" and
``selective" RF fields neglecting counter-rotating
components\cite{Abragam61}
are described by the Hamiltonians
\begin{align}
    H_{\rm RF}^{\rm B} &= -\hbar\Omega_{\rm RF}\sum_k
    \left[\cos(\omega t-\phi) I_{k}^{x} - \sin(\omega t-\phi)
    I_k^{y}\right],
    \label{eq:RF_B}\\
    H_{\rm RF}^{\rm S} &= -\sum_{l,k}\hbar\omega_{\rm RF}^l
    \left[\cos(\omega'_l t-\phi_l) I_{k}^{x} - \sin(\omega'_l t-\phi_l)
    I_k^{y}\right]
\label{eq:RF_S},
\end{align}
for broadband (B) and selective (S) pulses, where $I_k^{\pm} =
I_k^{x}\pm i I_k^{y}$. We denote the angular frequencies and
phases as $\omega$ and $\phi$ for a broadband pulse and
$\omega'_l$ and $\phi_l$ for a selective pulse targeted on spin
$l$.

When broadband pulses are applied, their time scale is
$1/n\delta\omega \ll 1/\delta\omega$. Thus the dipole coupling
between any pair of nuclear spins is well described by
Eq.~(\ref{eq:couplingz}). For simplicity, perfect delta-pulses are
assumed for broadband pulses; each pulse can be applied
instantaneously. In practice, the effect of finite pulse widths
due to limited RF power in a real NMR experiment\cite{Steffen00}
needs to be considered and accommodated before the proposed scheme
is implemented.  We make a further assumption requiring greater
scrutiny: we assume that $\emph{selective}$ pulses can be
performed quickly in comparison to the free dipolar dynamics, thus
allowing selective pulses that appear as $\delta$-functions with
respect to averages of the dipolar Hamiltonian. This requires that
$\delta\omega \gg D_{kl}$.  The applicability of these assumptions
will be discussed in Sec.~\ref{sec:realism}.

\section{Scheme for Decoupling and Recoupling}
\subsection{Broadband decoupling}
\label{sec:wahuha}

To establish our formalism for analyzing decoupling pulse
sequences, we summarize the celebrated multiple-pulse sequence for
decoupling, commonly called WHH-4 (or WAHUHA) \cite{WHH68}.

We begin by considering the effect of a broadband pulse on the
spins in the absence of selective pulses.  We consider the usual
``rotating reference frame," which rotates at the $RF$ frequency
$\omega$.  In this frame the spins undergo a time evolution
governed by the total Hamiltonian $H_{\rm Z} + H_{\rm RF}^{\rm B}
+ H_{\rm D}^z$, written in the form
\begin{equation}
    H = -\sum_k\hbar\delta\omega_k I_k^z
        - \frac{\hbar\Omega_{\rm RF}}{2}\sum_k\left[e^{-i\phi}I_k^{+} +
        e^{i\phi}I_k^{-}\right]+H_{\rm D}^z,
\end{equation}
where $\delta\omega_k
= \omega_k - \omega$. Since $\Omega_{\rm RF}\gg
|\delta\omega_k|\gg D_{kl}$, the broadband pulse causes a rotation
about the unit vector $\hat{n} = [\cos\phi, \sin\phi,0]$ by angle
$\theta=\Omega_{\rm RF}t$ a rotation we consider as instantaneous.
The unitary operator for such a rotation is $P_\phi=\exp(-i\theta
I^\phi)$.
%
The dipolar-decoupling sequence  WHH-4 consists of four such
broadband pulses, all with $\theta=\pi/2$.  For this sequence we
notate such a pulsed rotation with $\phi=0$ as $P_x$, with
$\phi=\pi/2$ as $P_y$, etc.

To understand the evolution of the spins during a fast sequence of
such pulses, we transform our Hamiltonian to a reference frame
that follows the pulses.  This reference frame is referred to as
the ``toggling frame."  A Hamiltonian $H$ written in this frame
will be denoted $\tilde{H}(t)$.  If there were no dipolar
evolution or field gradient, then spins in the toggling frame
would undergo no evolution whatsoever, even though the $P_\alpha$
pulses are causing rapid revolutions in the rotating frame.
Toggling-frame dynamics can be observed by ``stroboscopic"
observation, in which measurement with a particular phase
reference is performed only once per cycle, when the toggling
frame coincides with the rotating frame.

At the beginning of a cycle, the dipole coupling Hamiltonian of
Eq.~(\ref{eq:couplingz}) in the toggling frame coincides with
$H_{\rm D}^z$;
\begin{equation}
\tilde{H}_{\rm D}(0<t<\tau) = H_{\rm D}^z .
\end{equation}
When we apply a $P_x$ pulse at $t=\tau$ and a $P_{\bar{y}}$ pulse
at $t=2\tau$, then the toggling frame Hamiltonian becomes
\begin{multline}\label{eq:couplingy}
\tilde{H}_{\rm D}(\tau<t<2\tau) = P_{\bar{x}} H_{\rm D}^z P_x\\
=D_{kl}(2I_{k}^yI_{l}^y-I_{k}^xI_{l}^x-I_k^zI_l^z) \equiv H_{\rm
D}^y,
\end{multline}
and
\begin{equation}
\tilde{H}_{\rm D}(2\tau<t<3\tau) = P_{\bar{x}} P_y H_{\rm D}^z
P_{\bar y}P_x=H_{\rm D}^x.
\end{equation}
Similarly, a $P_y$ pulse at $t=4\tau$ returns the toggling frame
Hamiltonian to
\begin{equation}\label{eq:couplingx}
\tilde{H}_{\rm D}(4\tau<t<5\tau) = P_{\bar{x}} P_y
P_{\bar{y}}H_{\rm D}^z P_y P_{\bar y}P_x=H_{\rm D}^{y}.
\end{equation}
Finally a  $P_{\bar{y}}$ pulse at $t=5\tau$ brings the toggling
frame Hamiltonian back to $H_{\rm D}^{z}$;
\begin{equation}
\tilde{H}_{\rm D}(5\tau<t<6\tau) = P_{\bar{x}} P_y P_{\bar{y}}P_y
H_{\rm D}^z P_{\bar{y}}P_y P_{\bar y}P_x=H_{\rm D}^{z}.
\end{equation}

\begin{figure*}
\includegraphics[scale=0.4]{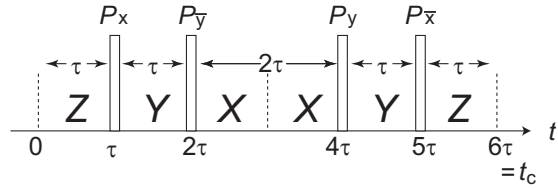}
\caption{\label{fig:WHH} Decoupling pulse sequence WHH-4.
\cite{Mehring83} $\bm{P}_{\alpha}$ and $\bm{P}_{\bar{\alpha}}$
stand for $\pi/2$ and $-\pi/2$ pulses around the $\alpha$-axis
($\alpha = x, y$). The system whose Hamiltonian is $H$ evolves in
the toggling frame by $Z=H$ for time 2$\tau$, $Y=P_{\bar{x}}H P_x$
for time 2$\tau$ and $X=P_{\bar{x}}P_y H P_{\bar{y}} P_x$ for time
2$\tau$ in one cycle $t_c = 6\tau$. In the case of the dipole
coupling Hamiltonian of Eq.~(\ref{eq:couplingz}), the zero-order
term and all odd-order terms in the average Hamiltonian vanish
over the period $t_c$.}
\end{figure*}

The decoupling pulse sequence WHH-4 shown in Fig.~\ref{fig:WHH} is
designed so that the system evolves in the toggling frame by
${H}_{\rm D}^x$ for time $2\tau$, ${H}_{\rm D}^y$ for time $2\tau$
and $H_{\rm D}^z$ for time $2\tau$ in one cycle.  The total cycle
time $t_c$ is $6\tau$.


We see that this particular WHH-4 cycle of form
$$\tau, P_x, \tau, P_{\bar{y}}, 2\tau,
P_y, \tau, P_{\bar x},\tau$$ causes a spin operator $I^z_j$ in the
rotating-frame Hamiltonian to go to $I^y_j$, then to $I^x_j$, and
then back again, in the toggling-frame Hamiltonian. We therefore
notate this cycle as $\mansfield{Z,Y,X}$ following the notation of
Mansfield\cite{mansfieldnotation}. Several equivalent WHH-4
sequences exist. For example, if $P_x$ is replaced by
$P_{\bar{x}}$ to make the sequence $\tau, P_{\bar{x}}, \tau,
P_{\bar{y}}, 2\tau, P_y, \tau, P_{x},\tau$, then we would find
$I^z$ going to $-I^x$, to $I^y$, and then back again, and so we
notate this cycle as $\mansfield{Z,\bar{Y},X}$.

In average Hamiltonian theory (AHT), the unitary evolution in the
toggling frame under this changing Hamiltonian is calculated by a
time ordered exponential, expanded via the Magnus
expansion\cite{Mehring83}.  The zero-order term is simply the
average Hamiltonian over the period of the sequence ${t_{\rm c}}$:
\begin{equation}
\bar{H}=\frac{1}{t_{\rm c}}\int_0^{t_{\rm c}} \tilde{H}(t) dt.
\end{equation}
For any WHH-4 sequence, the average Hamiltonian for the dipole
coupling is
\begin{equation}
    \bar{H}_{\rm D}=
    \frac{1}{6\tau}
    (2\tau H_{\rm D}^x + 2\tau H_{\rm D}^y + 2\tau H_{\rm D}^z) =
    0.
\end{equation}
It may be shown that all odd-order dipolar terms of the average
Hamiltonian vanish as well, due to the symmetry of the
cycle.\cite{Mehring83}
%

It is important to realize that dipolar-decoupling sequences have
been heavily developed in NMR over the past 40 years.  By
compounding these sequences with super-cycles, they can be made
robust against finite pulse width, pulse errors, and higher order
terms in AHT. More modern sequences such as BR-24\cite{br24} and
CORY-48\cite{cory48} have routinely proven their effectiveness in
reducing the timescale for dipolar evolution by three or more
orders of magnitude.

Among such pulse sequences is MREV-16, a variation of
MREV-8\cite{Mansfiled73,Rhim73}, one cycle of which is expressed
as
$\mansfield{Z,Y,X}\mansfield{Z,\bar{Y},X}\mansfield{Z,Y,\bar{X}}\mansfield{Z,\bar{Y},\bar{X}}.$
The cycle time is $t_c=24\tau$. This sequence is more robust
against pulse imperfections than WHH-4, but it also contains an
additional advantage.  The simple sequence $\mansfield{Z,Y,X}$
averages offset terms proportional to $I_k^z$, $I_k^x$ and $I_k^y$
to
\begin{equation}
    \bar{I}_k^z = \frac{I_k^x+I_k^y+I_k^z}{3}, \ \
    \bar{I}_k^x = \frac{2 I_k^x-I_k^y}{3}, \ \
    \bar{I}_k^y = \frac{I_k^y-2I_k^z}{3},
\label{zyxex}
\end{equation}
while the compound sequence MREV-16 yields the average operators
\begin{equation}
    \bar{I}_k^z = \frac{1}{3}{I}_k^z ,\qquad
    \bar{I}_k^x = \frac{2}{3}{I}_k^x ,\qquad
    \bar{I}_k^y = \frac{1}{3}{I}_k^y.
\end{equation}
In the case of MREV-16, $\bar{I}_k^x$ and $\bar{I}_k^y$ lie in the
plane perpendicular to $\bar{I}_k^z$, while this is not the case
for any WHH-4 sequence.
This property is critical for allowing selective RF rotations
during decoupling, as we discuss in the following section.


\subsection{Selective rotations}
\paragraph{Application of MREV-16 during selective pulses}
\label{sec:selectiverot} We now show how to produce selective spin
rotations on desired spins while decoupling the dipole coupling
between them by the MREV-16 broadband decoupling pulse sequence.
The result of such a selective RF pulse is to affect the evolution
of the desired spin ensemble, while leaving all other spins
unaffected.

Suppose we weakly irradiate the spins by selective pulses as in
Eq.~(\ref{eq:RF_S}), while \emph{simultaneously} applying the
broadband decoupling sequence MREV-16.
%
The RF Hamiltonian for selective pulses Eq.~(\ref{eq:RF_S}) is
written as
\begin{equation}
    H_{\rm RF}^{\rm S}= -\sum_l\frac{\hbar\omega_{\rm RF}^l}{2}
        \sum_k\left[e^{-i(\delta\omega'_l t+\phi_l)}I_k^{+} +
        e^{i(\delta\omega'_l t+\phi_l)} I_k^{-}\right],
    \label{eq:RF_S2}
\end{equation}
in the rotating frame at the angular frequency $\omega$ of the
broadband pulses, where $\delta\omega'_l = \omega'_l - \omega$.
Since $\Omega_{\rm RF}\gg |\delta\omega_l|$, Eq.~(\ref{eq:RF_S2})
varies slowly in time as compared to durations and intervals of
fast broadband pulses in the decoupling sequence. Therefore,
$H_{\rm RF}^{\rm S}$ is averaged by MREV-16 as if
Eq.~(\ref{eq:RF_S2}) were time-independent;
\begin{equation}
    \bar{H}_{\rm RF}^{\rm S} = -\sum_l\frac{\hbar\omega_{\rm RF}^l}{2}
    \sum_k\left[
        e^{-i(\delta\omega'_l t+\phi_l)}\frac{3I_k^{+}+I_k^{-}}{6}
    + {\rm h. c.}
    \right].
\end{equation}
Here we have made use of the fact that $I_k^{+}=I_k^x+i I_k^y$ is
averaged to $(3I_k^{+}+I_k^{-})/6$ by MREV-16.

We next show that this allows selective rotations on individual
spins in the presence of the Zeeman term, averaged to
\begin{equation}
    \bar{H}_{\rm Z} = -\sum_k \hbar\delta\omega_k\frac{I_k^z}{3}.
\end{equation}
Since $|\delta\omega_k|\gg |\omega_{\rm RF}^l|$ is assumed,
$\bar{H}_{\rm Z}$ dominates the time-evolution of the spins and
$\bar{H}_{\rm RF}^{\rm S}$ perturbs that evolution. It is
convenient to use the interaction picture,
\begin{equation}
    \bar{H}_{\rm RF}^{\rm S, I} (t) = e^{-i\bar{H}_{\rm Z}t/\hbar}
        \bar{H}_{\rm RF}  e^{i\bar{H}_{\rm Z}t/\hbar},
    \label{eq:RF_S_int}
\end{equation}
to find secular part of $\bar{H}_{\rm RF}$. Using the
transformation of $I_k^{\pm}$ into $I_k^{+}e^{-i\delta\omega_k
t/3}$ and $I_k^{-}e^{i\delta\omega_k t/3}$
in the interaction picture, one finds that the secular part of
Eq.~(\ref{eq:RF_S_int}) is
\begin{align}
 \bar{H}_{\rm RF}^{\rm S, I, sec} &= -\sum_l\frac{\hbar\omega_{\rm RF}^l}{4}
    \left[
        e^{-i\phi_l}I_l^{+} + e^{i\phi_l} I_l^{-}\right]\notag\\
        &=-\sum_l\frac{\hbar\omega_{\rm RF}^l}{2}\left[\cos\phi_l
        I_l^x+\sin\phi_l I_l^y
    \right],
\end{align}
where angular frequencies of the selective pulses $\omega'_l$ are
chosen to satisfy $\delta\omega'_l = \delta\omega_l/3$. This
Hamiltonian describes selective rotations of the target spins. The
simultaneous application of MREV-16 removes the dipole coupling
for all spins simultaneously: $\bar{H}_{\rm D}=0$.   Nonsecular
terms involving a resonant spin $(l)$ oscillate with frequency
$2\delta\omega_l/3$; nonsecular terms involving non-resonant spins
$(k\ne l)$ oscillate with frequencies $(\delta\omega_l\pm
\delta\omega_k)/3.$ This is similar to the principles of selective
rotations without broadband pulse sequences; errors on
non-resonant spins are neglected if the pulse bandwidth is
sufficiently small.   The influence of MREV-16 on this
consideration is that the bandwidth must be chosen 3 times smaller
for the same level of error.

\paragraph{Other decoupling sequences during selective pulses}
The choice of broadband sequence is not limited to MREV-16, but
for selective rotations only certain sequences will work.  To
discuss these constraints, let us generalize this result to an
arbitrary pulse sequence.  We still require the assumption that
the sequence is very fast in comparison to $\delta\omega_l'$, so
that Eq.~(\ref{eq:RF_S}) is averaged to a term of the form
\begin{multline}
\bar{H}_{\rm Z}+\bar{H}_{\rm RF}^{\rm S}(t)=
-\sum_k\hbar\delta\omega_k \vec\zeta\cdot\vec I_k
-\sum_{kl}\hbar\omega_{\rm RF}^l\times\\
\left[\cos(\omega_l't-\phi_l)\vec\xi\cdot\vec I_k-
\sin(\omega_l't-\phi_l)\vec\eta\cdot\vec I_k\right].
\label{genave}
\end{multline}
For example, Eq.~(\ref{zyxex}) shows that the sequence
$\mansfield{Z,Y,X}$ yields $\vec{\zeta}=(1,1,1)/3,$
$\vec{\xi}=(2,-1,0)/3$, and $\vec\eta=(0,1,-2)/3.$  The problem is
that $\vec\xi$ and $\vec\eta$ are not orthogonal to $\vec\zeta$.
Therefore Eq.~(\ref{genave}) contains terms of the form
$\cos(\omega_l't-\phi_l)(\hat{\vec\zeta}\cdot\vec\xi)\hat{\vec\zeta}\cdot\vec
I_k.$ and
$\sin(\omega_l't-\phi_l)(\hat{\vec\zeta}\cdot\vec\eta)\hat{\vec\zeta}\cdot\vec
I_k.$  In the interaction picture, these terms are diagonal in the
eigenbasis of $\bar{H}_{\rm Z}$; they do not represent transitions
between energy eigenstates, but rather a time-dependent
fluctuation of the energy eigenvalues.  As a result, simple
truncation of these terms leads to a poor approximation of the
dynamics.  It is very difficult to achieve a desired rotation
controlling only $\phi_l$ and $\omega_{\rm RF}^l$.

\begin{figure*}
\includegraphics[scale=0.4]{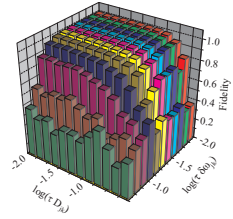}
\caption{\label{simulationfig} Fidelity versus
$\tau\delta\omega_{jk}$ and $\tau D_{jk}$.  The average wave
function overlap between dynamics calculated exactly and dynamics
calculated with zeroth order average Hamiltonian theory.  This
simulation averages over 3 random initial pure states, and uses
$\phi_j=0.7$, $\omega^j_{\rm RF}=\delta\omega_{jk}/100.$}
\end{figure*}

The situation with MREV-16 is much simpler, as
$\vec\zeta\cdot\vec\xi=\vec\zeta\cdot\vec\eta=0$ and so these
terms are not present.  The resulting average interaction
Hamiltonian is exactly the same as if there were no decoupling
pulse sequence, except that resonant offset frequencies are scaled
by the factor $\zeta=1/3$ and the Rabi frequency is scaled by the
factor $1/2.$  Arbitrary rotations can be achieved through simple
choice of $\phi_l$ and $\omega_{\rm RF}^l$, just as in normal NMR
techniques\footnote{This convenient aspect of MREV-16 is related
to another advantage, which is that under stroboscopic observation
in the rotating frame, the effective rotation angle for free spins
is parallel to the applied field, eliminating ``quad echo"
artifacts from measurement spectra at the cost of reducing
spectroscopic resolution. For this reason MREV-16 was employed in
Ref.~\onlinecite{Ladd04}.}.  Other, shorter sequences satisfy the
constraint $\vec\zeta\cdot\vec\xi=\vec\zeta\cdot\vec\eta=0$; for
example, the sequence
$\mansfield{Z,Y,X}\mansfield{Z,\bar{Y},\bar{X}}$ has
$\vec\zeta=(0,0,1)/3,$ $\vec\xi=(2,-1,0)/6$, and
$\vec\eta=(0,1,0)/3.$  The non-orthogonality of $\vec\xi$ and
$\vec\eta$ in this case, however, means that the relation between
the RF phase $\phi_l$ and the angle of rotation achieved by the
selective pulse becomes more complicated.  The general relation is
\begin{multline}
 \bar{H}_{\rm RF}^{\rm S, I, sec}=
 -\sum_l \frac{\hbar\omega_{\rm RF}^l}{2} \vec{I}_l\cdot\\
 \left[
    (\vec\xi-\hat{\vec{\zeta}}\times\vec\eta)\cos\phi_l
   +(\vec\eta-\hat{\vec{\zeta}}\times\vec\xi)\sin\phi_l
\right].
\end{multline}
So, for example, with sequence
$\mansfield{Z,Y,X}\mansfield{Z,\bar{Y},\bar{X}},$ we have
$$ \bar{H}_{\rm RF}^{\rm S, I, sec}=
-\sum_l \frac{\sqrt{17}}{12}\hbar\omega_{\rm RF}^l
[\cos(\phi_l+\phi_0)I^x+\sin(\phi_l-\phi_0)I^y],$$ where
$\phi_0=\tan^{-1}(1/4).$  A selective $\pi/2$ rotation of spin $l$
about $X$ may be achieved with $\delta\omega_l'=\delta\omega_l/3$,
$\phi_l=\phi_0,$ and $\omega_{\rm RF}^l t = 2\pi\sqrt{17}/5.$
\paragraph{Simulation of fast decoupling sequences}
We have tested the principles of such selective pulses in the
presence of fast decoupling techniques with a computer simulation.
In this simulation, two dipolar-coupled spins, labelled $j$ and
$k$, begin in a random pure state and are subjected to MREV-16.
Meanwhile, a selective $\pi$ pulse following the ideas above is
applied at one of the resonant frequencies $\delta\omega_j'$. The
time dynamics of this process are simulated in two ways. In the
first, the spin undergoes all the rotations of MREV-16 (treated as
perfect, zero-duration rotations) while also evolving according to
Eq.~(\ref{eq:RF_S2}). Here the evolution is found by direct
time-step integration of the Schr\"odinger equation.  In the
second simulation, the spin state at time $t$ is taken to be the
expectation from the average Hamiltonian:
\begin{equation}
\left|\psi(t)\right\rangle=\exp(-i\bar{H}_{\rm
Z}t/\hbar)\exp(-i\bar{H}_{\rm RF}^{\rm S, I, sec}t/\hbar)
\left|\psi(0)\right\rangle.
\end{equation}
The fidelity between the spin states calculated by these two
methods is calculated for different parameters $D_{jk}\tau$ and
$(\delta\omega_j-\delta\omega_k)\tau=\delta\omega_{jk}\tau.$  The
simulation is run several times, averaging over random initial
pure states. The resulting average fidelity is shown in
Fig.~\ref{simulationfig}.  We find that if $\tau$ is short enough
in comparison to $D_{jk}$ and $\delta\omega_{jk}$, the sequence
works extremely well, and that errors due to a dipolar coupling
which is too large or insufficient selectivity are independent, at
least at the two-spin level.


\subsection{Selective inversion for recoupling}
\label{sec:pheisen}

The combination of a broadband decoupling pulse sequence and
selective pulses allows independent single-qubit rotations, as
discussed above. The coupling between a pair of nuclear spins also
must be restored in order to perform two-qubit operations. We will
demonstrate how to selectively recover the coupling between a pair
of nuclear spins without affecting other nuclear spins.

We begin by applying rapid, broadband $\pi$-pulses that eliminate
the offset term $H_{\rm Z}$.  In what follows, these pulses are
always applied on the fastest possible timescale.  The average
evolution under these $\pi$-pulses is therefore governed by the
full dipolar term $\bar{H}_{\rm D}=H_{\rm D}$
[Eq.~(\ref{eq:couplingz})].  We then manipulate this dipolar term
with \emph{selective} $\pi$-pulses. These selective pulses may be
accomplished in the manner described in
Sec.~\ref{sec:selectiverot}, or by the usual soft NMR pulses if
the dipolar coupling is sufficiently small.  The broadband
$\pi$-pulses must be suspended during these pulses to allow
selectivity.

We now introduce a new ``second-toggled" reference frame in which
these selective rotations are treated as single, $\delta$-function
rotations. ``Free evolution" in the second-toggled frame
corresponds to the average evolution during rapid broadband $\pi$
pulses only and the dipolar Hamiltonian, which is unaffected by
those broadband $\pi$-pulses.

Working in this frame, we selectively rotate nucleus $k$ around
the $x$-axis by $\pi$ ($X_k$).
Then the second-toggled dipolar Hamiltonian is
\begin{equation}
    H_1^z \equiv X_k H_{\rm D}^{z} X_k =
    D_{kl}(-2I_{k}^{z}I_{l}^{z}-I_{k}^{x}I_{l}^{x}+I_{k}^{y}I_{l}^{y}).
    \label{eq:h_1}
\end{equation}
We further manipulate the second-toggled frame by applying
broadband $\pi/2$ pulses that correspond to the usual WHH-4
sequences, except spacing them by a time $T$ chosen to be much
longer than the selective pulse widths but much shorter than the
dipolar dynamics. Thus, as in Sec.~\ref{sec:wahuha}, we have
\begin{equation}
\begin{array}{l}
\secondtoggle{H}_{1}(\mbox{\hspace{1ex}}0^{+}<t<\mbox{\hspace{1ex}}T^{-}) = H_1^z,\\
\secondtoggle{H}_{1}(\mbox{\hspace{1ex}}T^{+}<t<2T^{-}) = P_{\bar{x}}H_1^z P_x,\\
\secondtoggle{H}_{1}(2T^{+}<t<4T^{-}) = P_{\bar{x}} P_y H_1^z P_{\bar{y}} P_x,\\
\secondtoggle{H}_{1}(4T^{+}<t<5T^{-}) = P_{\bar{x}}H_1^zP_x,\\
\secondtoggle{H}_{1}(5T^{+}<t<6T^{-}) = H_1^z,
\end{array}
\end{equation}
which averages to
\begin{align}
    \secondaverage{H}_1 &= \frac{1}{6T}\left(
    2T H^z_{1}
    + 2T P_{\bar{x}}H^z_1P_x
    + 2T P_{\bar{x}}P_y H^z_1 P_{\bar{y}} P_x \right)\nonumber\\
    &=
    -\frac{2}{3}D_{kl}^{\phantom{z}}(2I_{k}^{x}I_{l}^{x}+I_{k}^{y}I_{l}^{y}).
    \label{eq:bh_1}
\end{align}
We refer to this sequence as the $W_k(X)$ sequence.


Three more subcycles are required.  In the $W_k(Y)$ cycle, we
selectively rotate nucleus ${k}$ around the $y$-axis by $\pi$
($Y_k$).
The dipole coupling in the second-toggled frame is then
\begin{equation}
    H^z_2 \equiv Y_k H_{\rm D}^{z} Y_k =  D_{kl}^{\phantom{z}}(-2I_{k}^{z}I_{l}^{z}+I_{k}^{x}I_{l}^{x}-I_{k}^{y}I_{l}^{y}).
    \label{equ:h_2}
\end{equation}
Again taking this evolution through a WHH-4 cycle with
pulse-interval $T$, the second-averaged coupling averages to
\begin{align}
    \secondaverage{H}_2 &= \frac{1}{6T}\left(
    2T H^z_{2}
    + 2T P_{\bar{x}}H^z_2P_x
    + 2T P_{\bar{x}} P_y H^z_2 P_{\bar{y}} P_x \right)\nonumber\\
    &=
    -\frac{2}{3}D_{kl}^{\phantom{z}}(2I_{k}^{z}I_{l}^{z}+I_{k}^{y}I_{l}^{y}).
\end{align}
In the $W_k(Z)$ subcycle, we selectively rotate nucleus ${k}$
around the $z$-axis by $\pi$ ($Z_k$).

The double-toggled dipole coupling is
\begin{equation}
    H^z_3 \equiv
    D_{kl}^{\phantom{z}}(2I_{k}^{z}I_{l}^{z}+I_{k}^{x}I_{l}^{x}+I_{k}^{y}I_{l}^{y}),
    \label{equ:h_3}
\end{equation}
which averages to
\begin{align}
    \secondaverage{H_3} &= \frac{1}{6T}\left(
    2T H^z_{3}
    + 2T P_{\bar{x}}H^z_3P_x
    + 2T P_{\bar{x}}P_y H^z_3 P_{\bar{y}} P_x \right)\nonumber\\
    &=
    \frac{4}{3}D_{kl}^{\phantom{z}}(I_{k}^{x}I_{l}^{x}+I_{k}^{y}I_{l}^{y}+I_{k}^{z}I_{l}^{z})\nonumber\\
    &= \frac{4}{3}D_{kl}^{\phantom{z}}\Ib_{k}\cdot\Ib_{l}
\end{align}
over a slow WHH-4 cycle.  Finally, in the $W_k(-)$ subcycle, we do
not apply any selective pulses to nucleus $k$.  In this case, our
subcycle is just a fast WHH-4 where a $W_k(X)$ cycle would have
its selective $X_k$ pulse, combined with a slow WHH-4 cycle.  The
resulting second-averaged Hamiltonian is nothing more than the
dipolar coupling $H_{\rm D}^z$ averaged through a WHH-4 cycle,
which is of course zero.

In this way, three kinds of effective dipole coupling Hamiltonians
for a pair of nuclear spins are generated by selective inversion
of one or two of the nuclear spins under a slow WHH-4 sequence.
The important observation is that the sum of these three effective
Hamiltonians is zero:
\begin{equation}
\secondaverage{H_1}+\secondaverage{H_2}+\secondaverage{H_3}=0.
\end{equation}
This fact will be used to decouple unwanted couplings when the
coupling between a specific pair of nuclear spins is recovered.

\label{slowsequence} Suppose we apply the $W_k(\alpha)$ subcycle
to nucleus $k$ simultaneous with the $W_k(\beta)$ subcycle to
nucleus $l$ ($\alpha$, $\beta$ = $-$, $X$, $Y$, $Z$).  The
resulting second-averaged effective coupling Hamiltonian between
the two spins will be $\secondaverage{H}_1$,
$\secondaverage{H}_2$, or $\secondaverage{H}_3$ if
$\alpha\ne\beta$, or 0 if $\alpha=\beta$. Table~\ref{thetable}
summarizes all of the possibilities.
\begin{table}
\caption{Effective Hamiltonian generated by selective $\pi$-pulses
under a slow WHH-4 sequence. For example, when sequence $W(X)$
(See Fig.~\ref{fig:W}) is applied to $I_{k}$ and $W(-)$ to
$I_{l}$, the effective Hamiltonian over the cycle is
$\secondaverage{H}_{1}$, Eq.~(\ref{eq:bh_1}). When the same
sequence is applied to both spins, the dipole coupling is the same
as the natural dipole coupling Hamiltonian $H_{\rm D}^z$
[Eq.~(\ref{eq:couplingz})] and averages to zero over a slow WHH-4
cycle.\label{thetable}}
\begin{ruledtabular}
\begin{tabular}{l|cccc}
    & $W_l(-)$ & $W_l(X)$ & $W_l(Y)$ & $W_l(Z)$\\
    \hline
    $W_k(-)$ & 0 & $\secondaverage{H}_{1}$ & $\secondaverage{H}_2$ & $\secondaverage{H}_3$     \rule{0pt}{10pt}
\\
    $W_k(X)$ & $\secondaverage{H}_{1}$ & 0 & $\secondaverage{H}_{3}$ & $\secondaverage{H}_{2}$ \\
    $W_k(Y)$ & $\secondaverage{H}_{2}$ & $\secondaverage{H}_{3}$ & 0 & $\secondaverage{H}_{1}$\\
    $W_k(Z)$ & $\secondaverage{H}_{3}$ & $\secondaverage{H}_{2}$ & $\secondaverage{H}_{1}$ & 0\\
\end{tabular}
\end{ruledtabular}
\end{table}

\begin{figure*}
\includegraphics[scale=0.4]{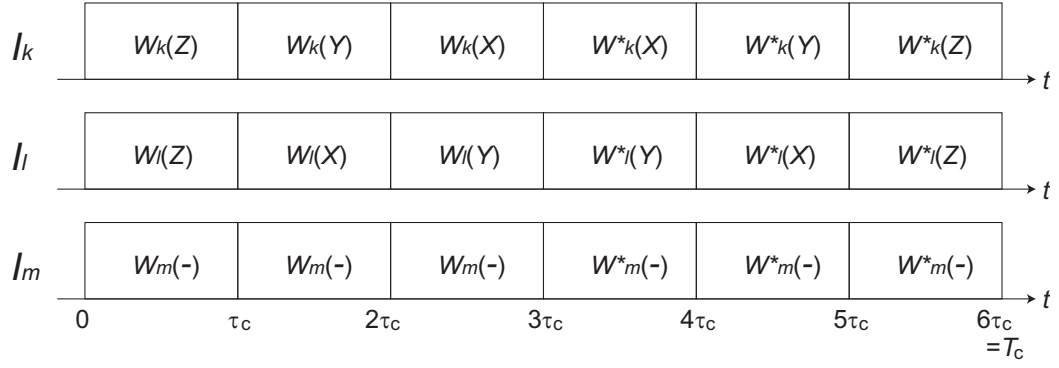}
\caption{\label{fig:super-WHH}Super-WHH sequence to couple nuclear
spins $I_{k}$ and $I_{l}$ while decoupling the rest of the spins
in a system $I_{m}$ from both. In the sequence
$W^{\ast}_k(\alpha)$, the phase and the order of the broadband
$\pi$-pulses are rearranged such that $W^{\ast}_k(\alpha)$ has a
time-reflection symmetry to $W_k({\alpha})$. Nuclear spins with
the same frequency are decoupled by this sequence as well. The
restored dipole coupling between $I_{k}$ and $I_{l}$ is $(8/9)
A_{kl}\Ib_k.\Ib_l$.}
\end{figure*}

Now we show how to recouple two spins $k$ and $l$ while the rest
of the spins, represented by $m$ in the following, are decoupled
from both spins $k$ and $l$.  Couplings between ensemble nuclei
(that is, nuclei with the same frequency) are also removed. Apply
$W_k(Z)$, $W_k(Y)$ and $W_k(X)$ sequentially to $I_{k}$ while
applying $W_m(-)$ three times to $I_{m}$, as shown in
Fig.~\ref{fig:super-WHH}. The effective coupling over this
supercycle, consisting of three slow WHH-4 cycles with cycle time
$\tau_{\text{c}}=6T$, is averaged to
\begin{equation}
    \frac{1}{3\tau_{\text{c}}}\left(\tau_{\text{c}}
    \secondaverage{H}_3
    +\tau_{\text{c}}\secondaverage{H}_2
    +\tau_{\text{c}}\secondaverage{H}_1\right)=0.
\end{equation}
The dipole coupling among $m$ spins is decoupled as well.

To spin $I_{l}$, apply $W_l(Z)$, $W_l(X)$ and $W_l(Y)$ in this
order. The dipole coupling between $I_{l}$ and $I_{m}$ is again
averaged to zero. The coupling between $I_{k}$ and $I_{l}$ is,
however,
\begin{equation}
    \frac{1}{3\tau_{\text{c}}}\left(0+\tau_{\text{c}}\secondaverage{H}_3
    +\tau_{\text{c}}\secondaverage{H}_{3}\right)
    = \frac{8}{9}D_{kl}\Ib_k.\Ib_l,
    \label{eq:recouple}
\end{equation}

In this way, the two spins $k$ and $l$ are recoupled according to
Eq.~(\ref{eq:recouple}) while the rest of nuclear spins remain
decoupled from spins $k$ and $l$ and from each other.

The present scheme works only in zeroth order in AHT. Dipolar
terms of order $n$ scale as $\tau_{\text{c}}^{n}D_{kl}^{n+1}$ for
other spins in the system with couplings as large as $D_{kl},$
including spins distant from $k$ and $l$ which are supposed to be
decoupled. To compensate for this, we must make
$\tau_{\text{c}}\ll D_{kl}^{-1}$ and perform quantum logic over
many cycles.  The resulting phase error for decoupled qubits over
a logic gate lasting $t\sim D_{kl}^{-1}$ scales as
$(\tau_{\text{c}}D_{kl})^n$. In the perfect-pulse limit, all odd
order terms can be eliminating by simply symmetrizing the
sequence, so that spin $k$ sees
$$
    W_k(Z) \ - \
    W_k(Y) \ - \
    W_k(X) \ - \
    W^{\ast}_k(X) \ - \
    W^{\ast}_k(Y) \ - \
    W^{\ast}_k(Z).
$$
and spin $l$ sees
$$
    W_k(Z) \ - \
    W_k(X) \ - \
    W_k(Y) \ - \
    W^{\ast}_k(Y) \ - \
    W^{\ast}_k(X) \ - \
    W^{\ast}_k(Z).
$$
In the sequence $W^{\ast}_k(\alpha)$, the phases and the orders of
the selective pulses are re-arranged such that
$W^{\ast}_k(\alpha)$ is time symmetrical to $W_k(\alpha)$. We
refer to this sequence as ``super-WHH." Its cycle time is
$T_{\text{c}}=6\tau_{\text{c}}=36T$.

\section{Feasibility of the Proposed Scheme}
\label{sec:realism}

Although the present scheme does not suffer from the $O(n)$
scaling of previous constructions,\cite{Leung00} it can only work
if the various timescales of the problem are well separated and
high power pulses are available.  In this section we discuss the
feasibility of these assumptions.

The approximations we used are summarized in
Table~\ref{approxtable}.
\begin{table}
\caption{Timescales.  It is assumed in the present scheme that
each timescale is much faster than the one below it.  Expected
orders of magnitude for these timescales are shown in the right
column. \label{approxtable}}
\begin{ruledtabular}
\begin{tabular}{rp{2.5in}l}
   Rate & Description & rad/s\\\hline
$\omega_0$ & Average Larmor frequency\rule{0pt}{2.5ex} & $10^9$\\
$\Omega_{\rm RF}$ & Rabi frequency for broadband pulses & $10^8$\\
$n\delta\omega$ & Number of qubits in gradient times
gradient-induced
frequency separation & $10^5n$\\
$\delta\omega$ & Gradient-induced frequency separation & $10^5$  \\
$\omega_{\rm RF}$ & Rabi frequency for selective pulses & $10^3$
\\
$T_{\text{c}}^{-1}$ & Inverse of super-WHH cycle time & 10\\
$D_{kl}$ & Near-neighbor dipolar coupling constant & 1 \\
$D_{\rm iso}$ & Intra-ensemble dipolar coupling constant &  0.01
\end{tabular}
\end{ruledtabular}
\end{table}
We now discuss the approximations made, the physical constraints
behind them, the error created, and potential improvements.

\subsection{High DC field: $\omega_0\gg \Omega_{\rm RF} \gg n\delta\omega$}

Laboratory magnetic fields rarely exceed 10-20~T due to the finite
critical currents in existing superconductors.  Obviously, it is
beneficial to have as large an external field as is available for
this scheme, as the Larmor frequency sets the ultimate physical
timescale for the computation rate.  If some qubits see an
external field comparable to $\Omega_{\rm RF}$, then nonsecular
terms become important and Bloch-Siegert shifts must be
considered.  Phase errors due to Bloch-Siegert shifts scale as
$(\Omega_{\rm RF}/\omega_0)^2$.

\subsection{High RF power: $\Omega_{\rm RF} \gg n\delta\omega \gg \delta\omega$}

Although the present scheme does not slow down as qubits are
added, maintaining the availability of broadband pulses becomes
more difficult as the number of qubits gets large.  However, the
amount of RF power available is limited only by such issues as
heat dissipation and arcing.  With careful engineering and the use
of microcoils,\cite{microcoils} Rabi frequencies approaching the
Larmor frequency are attainable.  Phase errors due to
insufficiently broadband pulses scale for the most off-resonant
nuclei as $(n\delta\omega/\Omega_{\rm RF})^2$.

Since $\Omega_{\rm RF}$ is ultimately limited by the Larmor
frequency $\omega_0$, it would seem that the present scheme puts
an upper limit on $n$, at least for reasonable clock speeds.
Further, if the RF power remains constant and the dipole coupling
is weakened as $n$ is increased, an $O(n)$ slowdown is again
incurred. However, even if finite power is assumed, $n$ need not
be the total number of qubits in the system.  One could imagine
instead a field gradient of finite extent, in which only a subset
$n$ in a much larger set of qubits $N$ is well-distinguished by
the field gradient, while all $N$ lie within a finite bandwidth
accessible by the available RF power.  The entire register can be
accessed by moving the gradient, which could be done by switching
of different gradient coils (as in magnetic resonance imaging), by
mechanical motion of ferromagnets (as in magnetic resonance force
microscopy), or by more exotic means involving manipulation of
local electronic hyperfine fields. By limiting a moveable gradient
to a small subsection of the qubit register in this way,
sufficiently broadband pulses can be applied with finite power;
the qubits poorly distinguished by weaker gradients are still
decoupled by the present scheme.  It is important to realize that
such concerns will only be important in very large quantum
computers with thousands of qubits.

\subsection{High field gradients or chemical shifts: $\delta\omega \gg \omega_{\rm
RF}\gg T_{\text{c}}^{-1}$}

Large gradients or chemical shifts are needed so that selective
pulses do not become prohibitively long.  Very large field
gradients are attainable by a variety of means. Gradients as large
as $\sim$10~T/$\mu$m with sufficient field homogeneity for
sustaining measurable qubit ensembles can be made with
ferromagnets.\cite{goldman00}  The qubit-qubit frequency
separation $\delta\omega$ can be made higher than $10$~kHz by
suitably separating the qubits.  Errors due to selective pulses
affecting nearby nuclei scale as
$\sinc(\pi\delta\omega/2\omega_{\rm RF})$. Compensation schemes
for such errors are possible.\cite{Steffen00}

Although an ideal symmetric sequence has a vanishing first order
term, a first order term does arrive in the presence of pulse
errors and finite pulse width.   For the simple sequences proposed
here, the first order terms linear in the pulse width can only be
neglected when the pulse widths of the selective pulses are much
shorter than the cycle time.   First order correction terms scale
linearly with $1/\omega_{\rm RF}T_{\text{c}}$.   It is important
to note, however, that compensation for finite pulse widths in
multiple pulse sequences can be achieved to high order by various
means.\cite{Mehring83}

\subsection{Slow dipole couplings: $T_{\text{c}}^{-1} \gg D_{kl} \gg D_{\text{iso}}$}

The timescale limitations discussed above limit the dipole
coupling strengths which can be manipulated by this sequence.  As
discussed in Sec.~\ref{slowsequence}, $n^{\text{th}}$ order
residual terms in the super-WHH sequence result in phase errors of
order $(D_{kl}T_{\text{c}})^n$.  Higher-order sequences can
alleviate this restriction, but in the presence of non-idealities
it is challenging to eliminate all second order terms.  For the
numbers used in Table~\ref{approxtable}, the dipole coupling must
then be slower than 1~Hz, which is possible for reasonably
separated nuclei, especially if $\gamma$ is low. Unfortunately,
this results in rather slow clock times. Although coherence times
can be extremely long,\cite{Ladd04} the exponential speedup of
quantum algorithms can only compensate for this large, constant
slow-down over classical computers for very large problem sizes.
Previous proposals for NMR computers\cite{Leung00} are no faster,
as the clock speed for such computers with several hundred qubits
would also fall slower than a hertz.  We re-emphasize the
important advantage to the present scheme: there is no adverse
scaling with $n$ with the super-WHH sequence.  Therefore, if the
clock cycle can be improved with higher magnetic fields, larger
gradients, and error compensation schemes, the scheme will remain
scalable for large quantum computers.

The final issue not resolved by the present scheme has been
discussed elsewhere.\cite{Ladd02} Copies of qubits in a single
ensemble are decoupled by the compound WHH cycles of the present
scheme.  However, when different qubits $k$ and $l$ are coupled,
all of the spins in ensemble $k$ are coupled to all of the spins
in ensemble $l$, leading to cross-couplings which ultimately
couple ensemble members.  This may only be compensated for in the
present scheme and in existing proposals by spatially separating
ensemble members in order to ensure that qubit couplings are much
stronger than couplings within each ensemble. This limits
available couplings to near-neighbor couplings only.\cite{Ladd02}

\section{Conclusion}
We have seen that the judicious use of hard and selective pulses
allows rotations of single-spin ensembles and selective couplings
between ensembles.  Single-spin rotations are performed by
application of a weak RF field in conjunction with a rapid
broadband decoupling sequence. Couplings between ensembles are
accomplished by altering slower decoupling sequences with
selective rotations, and constructing supercycles of such altered
subsequences which couple only desired nuclei. The feasibility of
this approach depends on the selection of a physical system that
exhibits a hierarchy of time scales.  These time scales must be
widely separated to minimize the phase error of hard pulses,
ensure the selectivity of soft pulses, minimize the impact of
finite pulse widths, abate higher-order contributions to the
average Hamiltonian, and mitigate inter-ensemble cross-couplings.
In solid-state quantum computer architectures where these
time-scale constraints are satisfied, the present scheme resolves
two previously unresolved issues, improving the temporal
scalability of such architectures: the linear slowdown of logic
operations; and the complication of added intra-ensemble
couplings.

\begin{acknowledgements}
This work was supported by the DARPA QuIST program and the ICORP
Quantum Entanglement Project, JST.  T.D.L. was supported by the
Fannie and John Hertz Foundation.
\end{acknowledgements}

\bibliography{decouplingplus-fy}

\begin{figure*}
\includegraphics[scale=0.4]{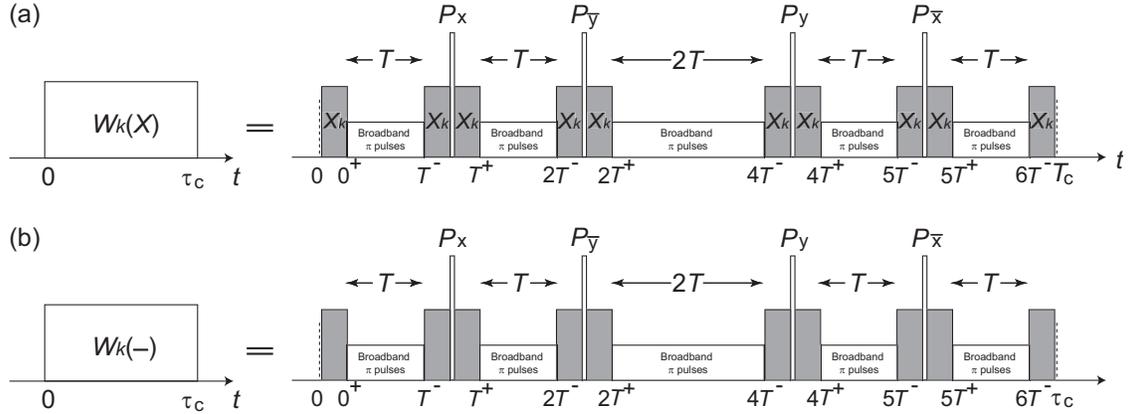}
\caption{\label{fig:W}(a) Sequence $W_k(X)$: Within each
time-interval of a slow broadband WHH-4 sequence (consisting of
$\pi/2$-pulses $P_x$, $P_{\bar{x}}$, $P_{y}$ and $P_{\bar{y}}$),
selective $\pi$-pulses around the $x$-axis (represented by $X_k$)
are inserted at the beginning and at the end of the interval.
Between the two selective $\pi$-pulses in an interval, fast
broadband $\pi$-pulses are constantly applied at a speed faster
than $1/\delta\omega$ in order to eliminate the Zeeman term.  The
selective $\pi$ pulses are produced by the combination of a
broadband decoupling sequence and selective pulses. Sequences
$W_k(Y)$ and $W_k(Z)$ are obtained by replacing $\pi$-pulses
around the $x$-axis ($X_k$) by ones about the $y$-axis ($Y_k$) and
the $z$-axis ($Z_k$), respectively. (b) Sequence $W(-)$: No
selective $\pi$-pulses are applied. The fast broadband
$\pi$-pulses and a broadband decoupling sequence are synchronized
with the sequences $W_k(X)$, $W_k(Y)$ or $W_k(Z)$ that are applied
to other spins.}
\end{figure*}

\end{document}